\documentclass[prl,draft,showpacs,twocolumn]{revtex4}
\usepackage{amsmath,amsfonts,amsthm}
\def\N{\mathbb N}
\def\E{\mathbb E}   %%% expectation
\def\qed{$\Box$}
\def\sig{ \sigma}
\newcommand{\eq}[1]{eq.~(\ref{#1})}  
\newcommand{\Ev}[1]{\E \left( #1 \right)}  
\newcommand{\Av}[1]{{\bf  Av } \left( #1 \right)}  

\newcommand{\Ert}[1]{\E^{(2)}_t \left( #1 \right)}
\def\K{{\kappa}}

\begin{document}
% FOR TWO COLUMN  ACTIVATE THE LINE BELOW
%@twocolumnfalse\endcsname
%%%
%%% \draft is now invoked as an option
%%%
%\draft
\title{An Extended Variational Principle for the
SK Spin-Glass Model}

\author{Michael Aizenman, Robert Sims, and Shannon L. Starr}
%%%
%%% Changed \address to \affiliation
%%%
\affiliation{
Departments of Physics and Mathematics, Jadwin Hall, Princeton
University, Princeton, NJ 08544}
\date{\today}
\begin{abstract}
The recent proof by F. Guerra that the Parisi ansatz provides a 
lower bound on the free energy of the SK spin-glass model 
could have been taken as offering some support to the 
validity of the purported solution.
In this work we present a broader variational principle, in which
the lower bound, as well as the actual value, are expressed through  
an optimization procedure for which ultrametic/hierarchal structures 
form only a subset of the variational class.  
The validity of Parisi's  ansatz for the SK model is still in question.  
The new variational principle may be of help in critical review of the issue.
\end{abstract}
%%%%%%%
\maketitle
%%%%%%%
\noindent{\em Introduction \qquad } 
The statistical mechanics of
spin-glass models is characterized by the existence of a diverse
collection of competing states, very slow relaxation of the quenched
dynamics, and a rather involved picture of the equilibrium state.

A great deal of insight on the subject has been produced through
the study of the Sherrington Kirkpatrick (SK) model~\cite{SK}.  After some 
initial attempts, a solution was proposed by G. Parisi which has the 
requisite stability and many other attractive features~\cite{P}.  Its
development has yielded a plethora of applications of the method, 
in which a key structural assumption is a particular form
of the replica symmetry-breaking (i.e., the assumption of
``ultrametricity'', or the hierarchal structure, of the overlaps among
the observed spin configurations)~\cite{MPV}.

Yet to this day it was not established that this very
appealing proposal does indeed provide the equilibrium structure of
the SK model.  A recent breakthrough is the proof by
F.~Guerra~\cite{Gue} that the free energy provided by  Parisi's
purported solution is a rigorous lower bound for the SK free energy.

More completely, the result of Guerra is that for any value of the
order parameter, which within the assumed ansatz is a {\em function},
the Parisi functional provides a rigorous lower bound.  
Thus, this
relation is also valid for the maximizer which yields the Parisi
solution.

In this work we present a variational principle for the free energy of
the SK model which makes no use of a Parisi-type order parameter, and
which yields the result of Guerra as a particular implication.  More
explicitly, the new principle allows more varied bounds on the free
energy, for which there is no need to assume a hierarchal organization
of the Gibbs state (e.g., as expressed in the assumed ultrametricity
of the overlaps~\cite{MPV}).  Guerra's results follow when the
variational principle is tested against the Derrida-Ruelle hierarchal
probability cascade models (GREM)~\cite{Ru}.

This leads us to a question which is not new: is the ultrametricity
an inherent structue of the SK mean-field model, or is it only a
simplifiying assumption.  The new variational principle may provide a
tool for challenging tests of this issue.

\noindent{\em The model \qquad} The SK model concerns Ising-type spins,
$\sig = (\sigma_1,\dots,\sigma_N)$,
with an a-priori equi-distribution over the values
$\{ \pm 1\}$, and the random Hamiltonian
\begin{equation}
        H_{N}(\sig)\ = \  \frac{-1}{\sqrt{N} }
        \sum_{1\le i < j \le N} J_{ij}  \,  \sigma_{i}\sigma_{j}
        \ - \ h \sum_{i=1}^N    \sigma_{i}
\end{equation}
where $\{J_{ij}\}$ are independent normal Gaussian variables. 

Our analysis applies to a more general class of Hamiltonians which 
includes all the even ``$p$-spin'' models~\cite{Der,GT2}.  Namely:
\begin{equation}
        H_N(\sig)\ =
        \ -  K_N(\sig) \ - \ h \sum_{i=1}^N    \sigma_{i}
\end{equation}
with
\begin{equation} \label{eq:pspin}
        K_N(\sig)\ = \  \sqrt{ \frac{N}{2} } \,  
        \sum_{r=1}^\infty  \frac{a_r}{  N^{r/2}} \sum_{i_1,\dots,i_r=1}^N
        J_{i_1\dots i_r} \sigma_{i_1}\cdots\sigma_{i_r}
\end{equation}
where all the $\{J_{i_1,\dots,i_r}\}$ are independent normal Gaussian
variables (for convenience, the tensor is not assumed here to be 
symmetric), 
and $\sum_{r=1}^{\infty}  |a_r|^2 =1$.   
As in \cite{GT2}, our argument requires that the  function
$  f(q) \ = \ \sum_{r=1}^{\infty}  |a_r|^2 \, q^r  \, $
be convex on $[-1,1]$.

One may note that $K_N(\sig)$ form a family
of centered Gaussian variables with the covariance
\begin{equation}
        \Ev{ K_N(\sig)\,  K_N(\sig')\  \vert\
        \sig\, ,\, \sig'}
        \ = \ \frac{N}{2} \, f( q_{\sig,\sig'}) \, ,
        \label{eq:fq}
\end{equation}
which depends on the spin-spin overlap:
$q_{\sig, \sig'} = \frac{1}{N} \sum_j 
\sigma_j \sigma'_j$.
The standard SK model corresponds to $f(q)=q^2$.

The partition function, $Z$,  the quenched free energy, $F$, and
what we shall call here the pressure, $P$, are defined as
\begin{gather}
        Z_{N}(\beta,h) =
        \sum_{\sigma_1,\ldots, \sigma_N = \pm 1} e^{-\beta\
        H_{N}(\sig)}
        \label{eq:Z} \\
                P_{N}(\beta,h) \ = \ \frac{1}{N}
        \Ev{ \log Z_{N}(\beta,h) } \ = \  - \beta\  F_{N}(\beta,h)
\end{gather}
where $\Ev{-}$ is an average over the random couplings $\{J_{ij}\}$.  
The thermodynamic limit for the free energy, i.e., the existence
of  $\lim_{N\to \infty}P_{N}(\beta,h) = P(\beta,h)$, was recently
established by Guerra-Toninelli~\cite{GT} through a much-awaited
argument.

\medskip

\noindent{\em The Variational Principle \quad }
 Our variational expression for
$P(\beta,h)$ employs a setup which may at first appear strange, but is
natural from the cavity perspective, when one 
considers the change in the total free 
energy caused by the addition of $M$ spins to a much larger system of size $N$.  
The expression for $Z_{N+M}/Z_{N}$ simplifies in the limit $N\to \infty$, 
at fixed $M$.   In the following idealized definition one may  
regard the symbol $\alpha$ as representing the configuration of the
bulk.  The discretness seen in the definition ($\sum \xi_{\alpha}$)  
is just for the convenience of the formulation of the variational bounds, 
and not an
assumption on the Gibbs state, though such an assumption may well be
true.  (A more general formulation is possible, 
but not much is lost by restricting attention to the
``ROSt'' defined below.)

% \begin{defn}
\noindent {\bf Definition (Random Overlap Structures):} { A {\em
random overlap structure} (ROSt) consists of a probability space
$\{\Omega, \mu(d\omega) \} $ where for each $\omega$ there is
associated a system of weights $\{ \xi_{\alpha}(\omega)\}$ and an
``overlap kernel'' $\{ \tilde{q}_{\alpha, \alpha'}(\omega) \} $ such that, 
for
each $\omega \in \Omega$,
\begin{enumerate}
        \item[i.] $\sum_{\alpha}  \xi_{\alpha}(\omega) \  \le \ \infty $,
        \item[ii.]  the quadratic form corresponding to
        $\{ \tilde{q}_{\alpha, \alpha'}\}$ is positive definite,
        \item[iii.] $\tilde{q}_{\alpha, \alpha}=1$, for each 
$\alpha$, and hence (by the Schwarz inequality) also:
 $|\tilde{q}_{\alpha, \alpha'}|\le 1$ for all pairs $\{ \alpha, 
\alpha'\}$.
\end{enumerate}
    }
% \end{defn}

An important  class of ROSt's is provided by the Derrida-Ruelle 
probability cascade model which is formulated in ref.~\cite{Ru} 
(called there GREM).   

Without additional assumptions, one may associate to the points in 
any ROSt   two independent families of centered Gaussian variables
$\{\eta_{j,\alpha} \}_{j=1,2,\ldots}$ and $\{\K_{\alpha}\}$
with covariances (conditioned on the random configuration
of weights and overlaps)
\begin{gather} \label{eq:etacov}
        \Ev{\,  \eta_{j,\alpha} \eta_{j',\alpha'} \, |
        \,   
        \tilde{q}_{\alpha, \alpha'}
                \, }
        \  = \ \frac{1}{2} \delta_{j,j'} \  f'(\tilde{q}_{\alpha, 
\alpha'})  \, ,\\ \label{eq:kapcov}
        \Ev{\,   \K_{\alpha}  \K_{\alpha'} \, |
        \,   
        \tilde{q}_{\alpha, \alpha'}     
                \, }
        \ = \  \tilde{q}_{\alpha, \alpha'} f'(\tilde{q}_{\alpha, \alpha'}) - 
f(\tilde{q}_{\alpha, \alpha'}) \, .
\end{gather}
The existence of such processes requires positive-definiteness of the 
joint covariance, but that is evident from the following explicit 
construction  in the case that the $\alpha$'s are   
$N$-vectors, with $q_{\alpha,\alpha'}=\frac{1}{N}\sum_j 
\alpha_j,\alpha_{j}'$:  
\begin{equation}
    \eta_{j,\alpha} \ = \ \sqrt{\frac{N}{2}}  \sum_r \frac{\sqrt{r} 
    \, a_r}{N^{r/2}} 
  \sum   \, 
    \widetilde{J}_{j,i_1,\ldots,i_{r-1}} 
\alpha_{i_1}\cdots\alpha_{i_{r-1}}   
\end{equation}
where the second sum is over $i_1,\ldots,i_{r-1}$ which range from  
$1$ to $N$, and
\begin{equation}
    \K_{\alpha} \ = \    \sum_r  \frac{\sqrt{r-1}\, a_r}{N^{r/2}} 
  \sum_{i_1,\ldots,i_r=1}^{N}  \, 
    \widehat{J}_{i_1,\ldots,i_r} \alpha_{i_1}\cdots\alpha_{i_r}   
\end{equation}

We shall now denote by $\E(\cdot)$ the combined average,
which corresponds to integrating over three sources of randomness:
the SK random couplings $\{ J_{ij} \}$,
the random
overlap structure described by the measure $\mu(d\omega)$, and
the Gaussian variables $\{\K_{\alpha}\}$ and $\{ \eta_{j,\alpha} \}$.

Guided by the cavity picture,  we associate with each
ROSt the following quantity:
\begin{multline} \label{ew:G} 
        G_M(\beta,h; \mu) \ = \\
=  \frac{1}{M} \ \Ev{ \, \log \left(
\frac{\sum_{\alpha, \sig} \xi_{\alpha}
\, e^{\beta \sum_{j=1}^{M} (\eta_{j,\alpha} + h ) \sigma_{j} } }
{\sum_{\alpha} \xi_{\alpha} \, \, e^{\beta \sqrt{M/2}
\, \K_{\alpha} }        }\right) }    
\end{multline}
where $\sig = (\sigma_1,\ldots, \sigma_M) $

Our main result is:

% \begin{thm}
\noindent{\bf Theorem 1}
{\em {\em i.} For any finite $M$, 
        \begin{equation}
                P_{M}(\beta,h) \ \le \   \inf_{(\Omega, \mu)} \ 
          G_M(\beta,h; \mu) \   \le \ P_{U}(\beta,h) \, , 
\label{eq:AFN}
\end{equation}
where the infimum is over random overlap structures (ROSt's) 
and $P_{U}(\beta,h)$ denotes the free energy $\times(-\beta)  $
obtained through the Parisi ``ultrametric'' (or ``hierarchal'')
ansatz.    \\ 
{\em ii.} The infinite volume limit of  the free energy
satisfies:
\begin{equation}
\label{eq:BFN}
     P(\beta,h) \ = \ \lim_{M\to \infty}
     \inf_{(\Omega, \mu)} \ G_M(\beta,h; \mu) \,  \, .
\end{equation} 
}
% \end{thm}
% \begin{proof}

\noindent{\em Proof \quad } 
These results can be seen as
consisting of two separate parts:  lower and an upper bounds, 
which are derived by different arguments.

{\em i.}  The upper bound: the left inequality in \eq{eq:AFN},
employs an interpolation argument which is akin to that used in the
analysis of Guerra~\cite{Gue}, but which here is formulated  in
broader terms without invoking the ultrametric ansatz.   The second 
inequality in (\ref{eq:AFN}) holds since the Parisi calculation represents the 
restriction of the variation to the subset of hierarchal ROSt's. 

To derive the first inequality let us introduce a family of 
Hamiltonians  for a mixed sytem of $M$ spins
$\sig = (\sigma_1,\dots,\sigma_M)$ and the ROSt variables $\alpha$, 
with a parameter $0\le t\le 1$: 
\begin{multline}
  - H_{M}(\sig, \alpha; t)  = 
  \sqrt{1-t} \left( K_{M}(\sig) + \sqrt{\frac{M}{2}}
  \, \K_\alpha \right )\  + \\
  +\  \sqrt{t} \,  \sum_{j=1}^{M} \eta_{j,\alpha} \sigma_{j}  
  \ +\  h \sum_{j=1}^M  \sigma_j\, , 
\end{multline}
and let 
    \begin{equation}
R_M(\beta,h;t) \ = \   \frac{1}{M} \ \Ev{ \, \log \left(
\frac{\sum_{\alpha, \sig} \xi_{\alpha}
\, e^{ - \beta H_{M}(\sig, \alpha; t) } }
{\sum_{\alpha} \xi_{\alpha} \, \, e^{\beta \sqrt{M/2}
\, \K_{\alpha} } } }
        \right)\, .
\end{equation}
Then 
\begin{align}
R_M(\beta,h;0) &= P_M(\beta,h)\, ,     \label{eq:int0}  \\ 
     R_M(\beta,h;1) &= G_M(\beta,h; \mu)\, , 
     \label{eq:int1} 
\end{align} 
and we shall  show that $ \frac{d}{dt}R_M(\beta,h;t) \ \ge \ 0$.

We use the following notation for  replica averages over 
pairs of spin and ROSt variables.  For any 
$X=X(\sig,\alpha)$  and $Y=Y(\sig,\alpha;\sig',\alpha')$:  
\begin{eqnarray}
\E^{(1)}_{t}(X)  & = &\Ev{\ \sum_{\alpha,\sig} 
        \ w(\sig,\alpha;t)\    X \ }  \\  
\Ert{Y} \ 
       & = &  \  \Ev{\ \sum_{\alpha,\sig} 
           \sum_{\alpha',\sig'}\ w(\sig,\alpha;t)\
            w(\sig',\alpha';t) \ Y \ }    \nonumber  
\end{eqnarray}
with the ``Gibbs weights''  
\begin{equation}\E^2_t
w(\sig,\alpha;t) =  
\xi_{\alpha} e^{-\beta H_{M}(\sig,\alpha;t)} \Big/
  \sum_{\alpha,\sig} \xi_{\alpha} 
  e^{-\beta H_{M}(\sig,\alpha;t)} \,  .  
\end{equation}
  
We now have
\begin{equation}
     \frac{d}{dt}R_M(\beta,h;t) \ = 
  - \frac{\beta}{M} \E_t^{(1)}\left( \frac{d}{dt} H_M(\sig,\alpha;t) 
  \right) \, .
\end{equation}
The term $\frac{d}{dt} H_M(\sig,\alpha;t)$ includes Gaussian 
variables, and one may apply to it the
generalized Wick's formula (Gaussian integration-by-parts) for 
correlated Gaussian variables, $x_1,\dots,x_n$:
\begin{multline}
\Av { x_1\, \psi(x_1,\dots,x_n) } \ = \\
= \ \sum_{j=1}^n \Av { x_1 x_j  } \, 
 \Av {\frac{\partial \psi(x_1,\dots,x_n)}{\partial x_j} } \, .
\end{multline}
The result is (after an elementary calculation):
\begin{equation} 
  - \frac{\beta}{M} \E_t^{(1)}\left( \frac{d}{dt} H_M(\sig,\alpha;t) 
  \right)\ = 
  \  \frac{\beta^2}{4}\, \E_t^{(2)}\left( 
  \varphi \right) \,   
\end{equation}
with  
% ( $[\ldots]$ indicating a shortcut in the calculation)
\begin{multline}
      \varphi(\sig,\alpha;\sig',\alpha')\ =  \\  
%       \left[ = \  \frac{1}{M}  \frac{d}{dt} 
%         \Av{(H(\sig,\alpha;t) - H(\sig',\alpha';t))^2 \Big\vert
%         \tilde{q}_{\alpha,\alpha'}, q_{\sig,\sig' }} \right ] \\
      = \ [f(q_{\sig,\sig'}) - f(\tilde{q}_{\alpha,\alpha'})]
        - (q_{\sig,\sig'} - \tilde{q}_{\alpha,\alpha'})
        f'(\tilde{q}_{\alpha,\alpha'})   \, . 
\end{multline}
Therefore,  
\begin{eqnarray}
\frac{d}{dt}R_M(\beta,h;t) \ =   \ \frac{\beta^2}{4}  \ \times   \  
    \qquad \qquad  \qquad \qquad \qquad \qquad   \\ 
    \Ert{ [f(q_{\sig,\sig'}) - f(\tilde{q}_{\alpha,\alpha'})]
  - (q_{\sig,\sig'} - \tilde{q}_{\alpha,\alpha'})
  f'(\tilde{q}_{\alpha,\alpha'}) }\,     \ge  0    \nonumber  \, .
  \end{eqnarray}
The last inequality,  which is crucial for us, 
follows from the assumed convexity of $f$.
For the SK model, the above expression  simplifies to
     $\Ert{\, 
     (q_{\sig,\sig'}-\tilde{q}_{\alpha, \alpha'} )^{2} \, } $.

 Putting the  positivity of the derivative together with  
(\ref{eq:int0}) and (\ref{eq:int1}) 
clearly implies the first bound in (\ref{eq:AFN}).

As was noted earlier, a particular class of random overlap 
structures is provided by the Derrida-Ruelle probability 
cascade models (GREM) of~\cite{Ru}, which are parametrized 
by a monotone function $x: [0,1] \to [0,1]$.   These models have 
two nice features: {\em i.} the distribution of 
$\{\xi_{\alpha}\}$  is invariant, except for a deterministic scaling 
factor,  under the multiplication by random factors as in 
(\ref{ew:G}) (consequently the value of  $G_M(\ldots,\mu_{x(\cdot)})$ 
for such ROSt does not depend on $M$),   
{\em ii.}  quantities like $G_M(\ldots,\mu_{x(\cdot)})$ 
can be expressed as the boundary values of the solution of a 
certain differential equation,  which depends on $x(\cdot)$.  
Evaluated for such models $G_M(\ldots,\mu_{x(\cdot)})$ reproduces 
the Parisi functional  for each value of the order parameter 
$x(\cdot)$.  The Parisi solution is obtained by optimizing (taking 
the {\em inf}) over the order parameter $x(\cdot)$.  This relation 
gives rise to the second inequality in (\ref{eq:AFN}).

{\em ii.} To prove (\ref{eq:BFN}) we need to supplement 
the first inequality in (\ref{eq:AFN}) by an opposite bound. 

Our analysis is streamlined by continuity arguments, 
which are enabled by the following basic estimate (proven by
two elementary applications of the Jensen inequality). 

\noindent{\bf Lemma 2} {\em 
Let $Z(H)$ denote the partition function for a system with 
the Hamiltonian $H(\sigma)$, and let $U(\sigma)$ be, for 
each $\sigma$, a centered Gaussian variable which is 
independent of $H$. Then 
\begin{equation} \label{Pb}
   0 \ \le \  \E\left(\log \frac{Z(H+U)}{Z(H)} \right) \ \le \ 
   \frac{1}{2}\, \E(\, U^2 \, ) \, . 
\end{equation}
} 

Using the above, it suffices 
to derive our result for interactions with the sum over $r$, in 
(\ref{eq:pspin}), truncated at some finite value.

A convenient tool is provided by the superadditivity of
$Q_N \equiv N\, P_N$, which was established in the work of
Guerra-Toninelli~\cite{GT} and its extensions~\cite{GT2,CDGG}. 
The statement is that for the systems discussed here (and in fact 
a broader class) for each $M, N \in \N$
\begin{equation}
     \label{eq:superadd}
        Q_{M+N}(\beta,h) \geq
        Q_M(\beta,h) + Q_N(\beta,h)\, .
     \end{equation} 
The superadditivity was used in \cite{GT} to establish the existence 
of the limit $\lim_{N\to \infty} P_N$.  However, it has a 
further implication based on the following useful fact.

\noindent{\bf Lemma 3 }
{\em
For any superadditive sequence   $\{Q_N\}$
satisfying (\ref{eq:superadd}) the following limits exist and satisfy
\begin{equation}
\lim_{N \to \infty} \,  Q_N / N
        = \lim_{M \to \infty} \liminf_{N \to \infty}\, \,
        [Q_{M+N} - Q_N]/M \, .
\end{equation}
}

For our purposes, this yields: 
\begin{equation}
    \lim_{N \to \infty} P_N \ = \  
    \lim_{M \to \infty} \liminf_{N \to \infty}\, 
    \frac{1}{M} \, \E\left(\log \frac{Z_{N+M}}{Z_{N}} \right) \, .
\end{equation}

We now claim, based on an argument employing the
cavity picture, that for any $M$
\begin{multline}
\label{eq:Opp}
     \liminf_{N \to \infty}
     \frac{1}{M}  \E\left(\log \frac{Z_{N+M}}{Z_{N}} \right) 
     \ \ge \ \inf_{(\Omega,\mu)} G_M(\beta,h;\mu)   
     \, ,
\end{multline}
which would clearly imply (\ref{eq:BFN}). 
The reason for this inequality is that  when a block of 
$M$ spins is added to a much larger ``reservoir'' of $N$ spins,
the change in the  free energy is 
exactly in the form of (\ref{ew:G}) -- except for  corrections 
whose total contribution   to $G_M$ is of order $O(\frac{M}{N})$.   
% (due to the 
(The spin-spin couplings within the 
smaller block and the subleading terms from the change 
$N \mapsto (N+M)$ in (\ref{eq:pspin}).) 
Thus, the larger block of spins acts as a ROSt.  

To see that in detail, let us split the system of $M+N$ spins 
into $\sigma = ( \tilde{\sigma}, \alpha)$, with 
$\tilde{\sigma}= (\sigma_1,\dots,\sigma_M )$ and $\alpha =
(\sigma_{M+1},\dots,\sigma_{M+N} )$.   With this notation, the 
interaction decomposes into
\begin{equation}
     K_{M+N}(\sig) \ = \
	\widetilde{K}_{N}(\alpha)
	+ \sum_{j=1}^M  \widetilde{\eta}_{j,\alpha}  \, \sigma_j 
	+U(\tilde{\sigma}, \alpha)
\end{equation}
where: {\em i.} $\{\widetilde{K}_N(\alpha) \}$ consists of  the 
terms  of $K_{M+N}(\sig) $  which involve only spins in the larger block, 
{\em ii.}  the second summand includes all the terms which involve exactly  
one spin in the smaller block,  and {\em iii.} 
$U$ consists of the remaining terms of $K_{M+N}(\sig) $, including the 
spin-spin interactions within the smaller block.  

One should note that 
$\{\widetilde{K}_N(\alpha) \} \neq \{ K_N(\alpha) \}$  since, 
as a consequence of the addition of the smaller block, the terms 
in  $\{\widetilde{K}_N(\alpha) \}$ are weighted by powers of $(N+M)$ 
rather than $N$, as presented in (\ref{eq:pspin}).   By the law of 
addition of independent Gaussian variables, $ \{ K_N(\alpha) \}$ 
(which are of higher variance than $ \{ \widetilde{K}_N(\alpha) \}$ ) 
have the same distribution as the sum of independent variables
\begin{equation}
\left\{ \widetilde{K}_N( \alpha) + \sqrt{ \frac{M}{2}}\kappa_{\alpha} 
\right\} \, , 
\end{equation}
where $\{ \kappa_{\alpha} \}$ are  centered Gaussian 
variables independent of $\widetilde{K}_N( \alpha) $. 
Up to factors $[1+O( \frac{M}{N})]$, the  covariances of  
$\{\widetilde{\eta}_{j, \alpha} \}$ and $\{ \kappa_{\alpha} \}$
satisfy (\ref{eq:etacov}) and (\ref{eq:kapcov}), respectively, 
and 
\begin{equation} \label{Ub}
\frac{1}{M}\, \E ( U( \tilde{\sigma}, \alpha)^2 ) \le C \frac{M}{N} \, .
\end{equation}
Taking 
\begin{equation}
\xi_{\alpha} := \mbox{ exp} \left[ \beta \left( \widetilde{K}_N(\alpha)+ h
    \sum_{i=1}^{N}  \alpha_i \right) \right], 
\end{equation}
we find that (\ref{eq:Opp}) follows by directly substituting the above into
(\ref{ew:G}) (using  (\ref{Pb}) and (\ref{Ub})).
\qed

\noindent{\em  Discussion \/ }
At first glance, the recent result of \cite{Gue} may be
read as offering some support to the widely shared belief that the
Parisi ansatz has indeed provided the solution of the SK model.
However, we showed here that the Guerra bound is part of a broader
variational principle in which no reference is made to the
key assumption of \cite{P}  that  in the limit $N\to \infty$ 
the SK Gibbs state develops a hierarchal organization.  
The reasons for such an organization,
which is equivalently expressed in terms of
``ultrametricity'' in the overlaps  $q_{\sig,\sig'}$,
are not a-priori obvious.
(A step, approaching the issue from a dynamical
perspective, was taken in ref~\cite{RA}, but this result has 
yet to be extended to the interactive cavity evolution.)
Our result (\ref{eq:AFN})  raises the possibility
that perhaps some other organizing principles may lead to
even lower upper-bounds.   
This reinstates  
the question whether the ultrametricity
assumption, which has enabled the calculation of \cite{P}, 
is correct in the context of the SK-type models.

It should be emphasized, however, that the question is not whether
the SK model exhibits replica symmetry breaking at low temperatures.
That, as well as  many other aspects of the accepted 
picture, are supported by both intuition and by rigorous
results (\cite{ALR,FZ,PS,Tal,NS}).
The question concerns the validity of a solution-facilitating
ansatz about the hierarchal  form of the replica symmetry breaking.
The interest in this question is enhanced by the fact that this
assumption yields a computational tool
with many other applications~\cite{MPV}.

\noindent{\em Acknowledgements \/ } This work was supported in part 
by NSF grant PHY-9971149.    R. Sims and S. Starr gratefully 
acknowledge the support of NSF Postdoctoral Fellowships (MSPRF). 

%   \bibliography{SK}  % to rerun bibtex 
%  \bibliographystyle{apsrev}
%   \end{document} % comment out when done

\end{document}